\documentclass{aa}

\usepackage{graphicx}
\usepackage{txfonts}
\usepackage{natbib}

\bibpunct{(}{)}{;}{a}{}{,} % to follow the A&A style

\newcommand{\matrixtt}[4]{\left( \begin{array}{cc}#1&#2\\#3&#4\\\end{array} \right)}

\newcommand{\herm}{H}
\newcommand{\jones}[2]{\vec {#1}_{#2}}
\newcommand{\jonesinv}[2]{\vec {#1}^{-1}_{#2}}
\newcommand{\jonesT}[2]{\vec {#1}^{\herm}_{#2}}
\newcommand{\jonesTinv}[2]{\vec {#1}^{{\herm}-1}_{#2}}
\newcommand{\coh}[2]{\mathsf{{#1}}_{{#2}}}

\newcommand{\EDIT}[1]{#1}

\begin{document}

\title{Revisiting the radio interferometer measurement equation.\\II. Calibration and direction-dependent effects}

\author{O.M.\ Smirnov}

\institute{Netherlands Institute for Radio Astronomy (ASTRON)\\
  P.O. Box 2, 7990AA Dwingeloo, The Netherlands \\
  \email{smirnov@astron.nl}}

\date{Received 5 Nov 2010 / Accepted 5 Jan 2011}

\titlerunning{Revisiting the RIME. II. Calibration and DDEs.}
\authorrunning{O.M.\ Smirnov}

\abstract%
%optional context
{Paper I of the series re-derived the radio interferometry measurement equation (RIME) from first principles, and extended the Jones
formalism to the full-sky case, incorporating direction-dependent effects (DDEs).}
%aims
{This paper aims to describe both classical radio interferometric calibration (selfcal and related methods), and the
recent developments in the treatment of DDEs, using the RIME-based mathematical framework developed in Paper I. It also
aims to demonstrate the ease with which the various effects can be described and understood.}
%methods
{The first section of this paper uses the RIME formalism to describe self-calibration, both with 
a full RIME, and with the approximate equations of older software packages, and shows how this is affected 
by DDEs. The second section gives an overview of real-life DDEs and proposed methods of dealing with them.}
%results
{A formal RIME-based description and comparison of existing and proposed approaches to the problem of DDEs.}
%optional conclusions
{}


\maketitle

\section*{Introduction}

\EDIT{
Paper I of this series \citep{RRIME1} extended the RIME formalism \citep{ME1,ME4} to the full-sky case, culminating in the following equation for the visibility matrix measured by interferometer $pq$:
 
\begin{eqnarray}\label{eq:me-allsky}
\coh{V}{pq} & = & \jones{G}{p} \left( \iint\limits_{lm} \coh{B}{pq} \mathrm{e} ^{-2\pi i(u_{pq} l+v_{pq} m)} \,dl\,dm \right) \jonesT{G}{q}, \\
\nonumber \coh{B}{pq} & = & \jones{E}{p} \coh{B}{} \jones{E}{q}
\end{eqnarray}

The $\coh{B}{}$ term is a $2\times2$ {\em brightness matrix}, describing the polarized sky brightness as a function of direction $l,m$. The $\jones{G}{p}$ Jones matrices represent the per-antenna direction-independent effects (DIEs), which are the provenance of traditional second-generation calibration (2GC) techniques, most notably selfcal. The $\jones{E}{p}$ Jones matrices represent the direction-dependent effects (DDEs). 

DDEs violate the traditional premise of 2GC, which is that an interferometer array measures the Fourier transform of one ``common'' sky. Instead, in the presence of DDEs, each baseline sees its \emph{own} apparent sky $\coh{B}{pq}$. The traditional premise only holds when the DDEs are \emph{identical} across all antennas, and constant in time: $\jones{E}{p} \equiv \jones{E}{}$. Under this condition, the apparent sky becomes the same on all baselines: ($\coh{B}{\mathrm{app}} =  \jones{E}{} \coh{B}{} \jonesT{E}{}$), and the full-sky RIME becomes simply: 

\begin{equation}\label{eq:me-allsky-simple}
\coh{V}{pq} = \jones{G}{p} \coh{X}{pq} \jonesT{G}{q},
\end{equation}

where $\coh{X}{pq} = \coh{X}{}(u_{pq},v_{pq})$, and the matrix function $\coh{X}{}(u,v)$, called the {\em sky coherency}, is the (element-by-element) two-dimensional Fourier transform of the matrix function $\coh{B}{\mathrm{app}}(l,m)$.

Section~\ref{sec:calibration} of this paper reviews the 2GC calibration problem, and shows how the RIME formalism can be (and has been) applied. Section~\ref{sec:ddes} then looks at the problem of DDEs, describes how they impact calibration, and discusses some current and future approaches.
}

\section{\label{sec:calibration}Calibration and the RIME}

In the traditional (2GC) view, \emph{calibration} refers to a process by which the instrumental errors are estimated and corrected for, \emph{imaging} is the processes of turning the corrected visibilities into an image, followed by \emph{deconvolution} to take out the effects of the point spread function. While algorithms such as Cotton-Schwab CLEAN \citep{Schwab:csclean} have blurred the boundaries between imaging and deconvolution, the separation between calibration and imaging is firmly entrenched in 2GC selfcal implementations (where the two processes are typically implemented via completely separate tools), and has historically led to a divergence of the algorithm development community into ``calibration people'' and ``imaging and deconvolution people''. 

The RIME, and recent developments in understanding of DDEs, have been eroding this distinction. On the one hand, advances in image reconstruction techniques \citep[for an overview, see][]{Rau:DDEs} have been usurping some traditional functions of calibration, while new methods of source modelling on the calibration side, such as the use of shapelets \citep{Yatawatta:shapelets}, rely on increasingly elaborate models being constructed for a large part of the flux (with traditional imaging then only used for the lower-level residuals). In RIME terms, both processes should be thought of as two aspects of the same optimization problem: estimating $\coh{B}{}(l,m)$, $\jones{E}{p}(l,m)$ and $\jones{G}{p}$ in an equation such as (\ref{eq:me-allsky}) that yield the best fit to a set of observed visibilities  (``data'') $\coh{D}{pq}$. Traditional selfcal solves for the direction-independent terms $\jones{G}{p}$, traditional imaging yields the $\coh{B}{}(l,m)$, and the non-trivial DDEs $\jones{E}{p}(l,m)$ have (traditionally) been ignored. The historical calibration--imaging separation corresponds to a two-stage recursive optimization process. 

\subsection{\label{sec:implicit-me-newstar}Implicit RIMEs}

Existing 2GC packages all make use of some implicit version of the RIME. It is useful to consider at least one example in depth. In Paper III \citep{RRIME3}, I shall be comparing the results of a MeqTrees calibration using an explicit RIME to those obtained with the NEWSTAR package on the same data. NEWSTAR therefore makes for a perfect example.

The exact form of the RIME implemented by NEWSTAR depends on the options used.\footnote{The version of the NEWSTAR RIME covered here does not include bandpass or polarization calibration. These options \emph{are} available in NEWSTAR, but they were not used for the calibration described in Paper III \citep{RRIME3}.} The one relevant to the reductions of Paper III is:

\begin{eqnarray}\label{eq:newstar-rime}
\coh{V}{pq} & = & \jones{G}{p} \left ( \coh{M}{pq} \ast \coh{X}{pq} \right ) \jonesT{G}{q}, \\
\nonumber \coh{X}{pq} & = & \sum_{s} E^2_s \coh{X}{spq} 
\end{eqnarray}

The constituent parts of this equation are as follows:
\begin{description}
\item[$\coh{X}{spq}$] is the coherency of source $s$. NEWSTAR sky models are composed of discrete point sources or extended Gaussian components. For a point source, $\coh{X}{spq} = K_p Q_{spq} \coh{B}{s} K^\herm_q$.

\item[$\coh{B}{s}$] is the source brightness. This can be further parametrized in terms of Stokes $IQUV$, spectral index and rotation measure.

\item[$Q_{spq}$] is a per-source correction factor to account for time and bandwidth smearing \citep[see Paper I,][Sect.~5.2]{RRIME1}.

\item[$E_s$] is the primary voltage beam gain. NEWSTAR uses an analytic approximation of the WSRT beam (see Sect.~\ref{sec:EJones:wsrt}). This is implicitly treated as a trivial DDE, i.e. constant in time and the same across all stations.

\item[$\coh{X}{pq}$] is thus the ``model visibilities'', i.e. the sum of coherencies of all sources in the model.

\item[$\jones{G}{p}$] is a {\em diagonal} matrix of complex per-station gain terms.

\item[$\coh{M}{pq}$] is a $2\times2$ matrix of multiplicative interferometer errors \citep[see Paper I,][Sect.~5.3]{RRIME1}, and ``$\ast$'' is element-by-element multiplication). Here it is on the inside of the equation rather than on the outside as in Eq.~(24) of Paper I: this is due to the way NEWSTAR uses ``corrected data'' in its selfcal procedure.%\footnote{In this particular case it makes no difference: since only the $XX$ and $YY$ correlations are used, all matrices in Eq.~(\ref{eq:newstar-rime}) are diagonal, and for diagonal matrices ``$\ast$'' is equivalent to matrix multiplication, and commutes. For the full-polarization case, the NEWSTAR treatment of $\coh{M}{pq}$ is incompatible with Eq.~(\ref{eq:me:closure-errors}).}

\end{description}

NEWSTAR's calibration and imaging procedure typically consists of some combination of and/or iteration over the following steps:

\begin{description}

\item[Gain calibration:] find $\jones{\tilde{G}}{p}$ that minimizes $|\jones{G}{p}\coh{X}{pq}\jonesT{G}{q} - \coh{D}{pq}|$ in a least-squares sense. Compute ``corrected data'' as $\jones{D}{pq}' = \jonesinv{\tilde{G}}{p} \coh{D}{pq} \jonesTinv{\tilde{G}}{q}.$

\item[Closure errors:] find $\coh{\tilde{M}}{pq}$ that minimizes $|\coh{M}{pq} \ast \coh{X}{pq} - \coh{D}{pq}'|$.
Compute ``corrected data'' as $\coh{D}{pq}'' = \coh{D}{pq}' \div \coh{\tilde{M}}{pq}$ (where ``$\div$'' is element-by-element division -- the inverse of ``$\ast$'').

\item[Model subtraction:] Compute ``residual data'' as $\coh{R}{pq} = \coh{D}{pq}'' - \coh{X}{pq}$. $\coh{R}{pq}$ thus contains the visibility contribution of faint background sources not present in the model, corrected for the estimated antenna gains and interferometer errors.

\item[Imaging and deconvolution:] turn the $\coh{R}{pq}$ visibilities into an image, and deconvolve it using H\"ogbom CLEAN. 

\item[Source finding:] Perform a source finding procedure on the residual image to update the sky model.

\item[Model update:] Solve for the parameters of the new sources by minimizing $|\coh{D}{pq}'' - \coh{X}{pq}|$ (usually on a small subset of the data).

\item[Model restore:] Add the sky model into residual images (after another calibration/subtraction cycle, if the model was updated), using a Gaussian restoring beam.

\end{description}

Calibration procedures implemented by other 2GC packages may differ in detail, but are very similar in principle. The crucial common concepts are: (a) the use of an equation such as (\ref{eq:newstar-rime}), which clearly separates the model visibilities ($\coh{X}{pq}$) from antenna-based errors ($\jones{G}{p}$), and (b) the procedure of \emph{correcting} visibilities (whether on-the-fly or in storage) by applying the inverse of the $\jones{G}{p}$ solutions. Both concepts break down when DDEs become involved, as will be discussed in Sect.~\ref{sec:ddes}.

\subsection{Explicit RIMEs}

An example of an explicit RIME is implemented in CASA. This also relies on the concept of model visibilities:

\begin{equation}\label{eq:casa-rime}
\coh{V}{pq}  =  \jones{J}{p} \coh{X}{pq} \jonesT{J}{q} \\
\end{equation}

Here, $\coh{X}{pq}$ is the model visibility (which may be computed from an image and/or
a list of NEWSTAR-like components), and $\jones{J}{p}$ is composed of several different Jones terms, typically \citep[Appendix E.1]{CASA:UserRef}:

\begin{equation}\label{eq:casa}
\jones{J}{p} = \jones{B}{p} \jones{G}{p} \jones{D}{p} \jones{E}{p} \jones{P}{p} \vec  T_p
\end{equation}

Each term has its own specific implementation (in case of known terms) and parametrization (in case of solvable terms). Finally, multiplicative interferometer-based errors ($\coh{M}{pq}$) may be optionally applied to either the outside of the equation \citep[as per Eq.~24 of Paper I,][]{RRIME1}, or to $\coh{X}{pq}$ itself (a-la NEWSTAR, see Eq.~\ref{eq:newstar-rime} above). 

Conceptually, calibration in CASA is very similar to the procedure described in the previous section, but the use of an explicit RIME confers several advantages. The known terms of the Jones chain (Eq.~\ref{eq:casa}) can be taken into account properly, while the solvable terms can be solved for in different combinations. The caveats of using such a specific form of the RIME have already been discussed in Paper I \citep[Sect.~6.2]{RRIME1}.

Note that although CASA also relies on the essentially 2GC-rooted concepts of model and corrected visibilities, the framework has been successfully used for the development of algorithms for calibration and correction of DDEs, namely \emph{W-projection} \citep{Cornwell:wproj}, \emph{pointing selfcal} \citep{SB:pointing} and \emph{AW-projection} \citep{SB:imageplane}. I will discuss these further in Sect.~\ref{sec:ddes}.

\subsection{\label{sec:phenomenological}Phenomenological RIMEs}

My experiments with calibration in MeqTrees have favoured \emph{phenomenological} RIMEs \citep{meqtrees}. Rather than writing out long Jones chains such as that of Eq.~(\ref{eq:casa}), which attempt to follow the physics of the signal propagation chain, the phenomenological approach consists of using a RIME with the minimum number of solvable terms needed to represent the cumulative effect of the chain. Each phenomenological term then ends up subsuming several different physical effects. The rationale for this approach is that, on the one hand, we only need to capture the overall effect for purposes of calibration, while on the other hand, the individual effects often cannot be distinguished at all, apart from their different behaviour in time and frequency -- which we try to capture with individual phenomenological terms.

For example, a full-polarization bandpass-gain calibration of the WSRT can be done\footnote{In the absence of DDEs.} using the following phenomenological RIME:

\[
\coh{V}{pq} = \jones{G}{p} \jones{B}{p} \coh{X}{pq} \jonesT{B}{q} \jonesT{G}{q} 
\]

Here, $\jones{G}{p}$ is a solvable diagonal complex matrix with rapid variation in time, and none in frequency. This subsumes antenna/receiver gains ($G$-Jones, in CASA nomenclature) and atmospheric phase ($T$-Jones). $\jones{B}{p}$ is a solvable full $2\times2$ complex matrix with high variability in frequency, but little to none in time. This subsumes bandpass ($B$-Jones), polarization leakage ($D$-Jones) and on-axis beam gain ($E$-Jones). More real-life examples of phenomenological RIMEs will be discussed in Paper III \citep{RRIME3}.

Where a specific Jones term is known from \emph{a priori} considerations, it can and should be inserted into a phenomenological RIME. For example, the equation above would not be suitable for polarization calibration of the VLA because of parallactic angle rotation. The equation would need to be rewritten with an extra $P$-Jones term, which is not solved for, but rather computed analytically:

\begin{equation}\label{eq:phenom-vla}
\coh{V}{pq} = \jones{G}{p} \jones{B}{p} \jones{P}{p} \coh{X}{pq} \jonesT{P}{q} \jonesT{G}{q} \jonesT{B}{q} 
\end{equation}

One must be mindful of matrix (non)commutation when constructing phenomenological RIMEs. The reason the full CASA Jones chain of Eq.~(\ref{eq:casa}) can be captured by the much simpler Eq. ~(\ref{eq:phenom-vla}) is because some Jones matrices \emph{do} commute \citep[see also Paper I,][Sect.~1.6]{RRIME1}. In particular, $T$-Jones is scalar and so commutes with everything, while the CASA $B$-Jones and $G$-Jones are diagonal and so commute among themselves. This allows us to rewrite Eq.~(\ref{eq:casa}) as:

\[
\jones{J}{p} = (\jones{G}{p}\jones{T}{p})(\jones{B}{p}\jones{D}{p}\jones{E}{p})\jones{P}{p},
\]

which makes the link to Eq.~(\ref{eq:phenom-vla}) obvious. 

To give a counter-example, in the presence of significant Faraday rotation (time-variable or differential, see Sect.~\ref{sec:DFR}), this equation is not appropriate, because the Faraday rotation term $\jones{F}{p}$ (placed at the right-hand side of the chain) does \emph{not} commute, and so would necessitate an extra term in Eq.~(\ref{eq:phenom-vla}). 

\subsection{The impact of the RIME on calibration}

The reasoning used above to construct phenomenological RIMEs illustrates one of the biggest benefits that the RIME formalism has brought to the field of calibration. Pre-RIME, descriptions of signal propagation effects were \emph{ad hoc} and approximate, while arguments about the order in which they should be calibrated for were difficult to follow. The RIME formalism has recast all this in terms of straightforward and rigorous matrix algebra.

The second benefit of the RIME formalism is the clarity it has brought to polarization calibration. Note that the implicit NEWSTAR RIME given above (Eq.~\ref{eq:newstar-rime}) ignores polarization effects almost completely. NEWSTAR does have some polarization calibration capabilities (as do other 2GC packages), but these are specifically tuned to the WSRT case. The RIME formalism allows for a much more general description of polarization effects. The $\jones{D}{}$ and $\jones{P}{}$ terms of the CASA RIME (Eq.~\ref{eq:casa}) are an example, but see also the discussion of differential Faraday rotation in Sect.~\ref{sec:DFR}. 

Perhaps most importantly, the RIME gives us the mathematical language to tackle the problem of DDEs, which will be the subject of the next section.

\subsection{Calibration ambiguities}

No discussion of calibration with the RIME can be complete without mentioning the {\em ambiguity} problem pointed out by \citet{ME4,ME5}. In classical selfcal, there is a well-known flux and position ambiguity: multiplying all the antenna gains by a complex factor $a$, and the source coherency by $a^{-2}$, does not change the observed visibilities. Therefore, selfcal by itself cannot determine absolute fluxes and positions -- these require known calibrators. There is a full-polarization equivalent to this, but it is extremely difficult to formulate and understand outside the RIME formalism.

For a direct analogue, consider a RIME such as that in Eq.~(\ref{eq:me-allsky-simple}). For any non-singular matrix $\jones{A}{}$, we have: 

\[
\coh{V}{pq} = \jones{G}{p} \coh{X}{pq} \jonesT{G}{q} = 
(\jones{G}{p} \jones{A}{})(\jonesinv{A}{} \coh{X}{pq} \jonesTinv{A}{})(\jones{G}{q} \jones{A}{})^H \]

In other words, we can multiply all the per-antenna uv-Jones terms by $\jones{A}{}$, and the source coherency by $\jonesinv{A}{}$ and $\jonesTinv{A}{}$, without changing the observed visibilities. Therefore, we need known calibrators to properly fix the $\jones{G}{p}$'s. Having observed a calibrator source, we can fix the brightness $\coh{B}{}$ (and therefore the coherency $\coh{X}{pq}=K_p\coh{B}{}K_q$), and solve for $\jones{G}{p}$. However, it is easy to see that an unpolarized calibrator alone is insufficient. The brightness (and coherency) matrix of an unpolarized source is scalar, so for any {\em unitary}\footnote{$\jones{U}{}$ is {\em unitary} if $\jones{U}{}\jonesT{U}{}=1$.} matrix $\jones{U}{}$, we have $\jones{U}{}\coh{X}{pq}\jonesT{U}{}=\jones{U}{}\jonesT{U}{}\coh{X}{pq} = \coh{X}{pq}$, or: 

\[
\coh{V}{pq} = \jones{G}{p} \coh{X}{pq} \jonesT{G}{q} = (\jones{G}{p} \jones{U}{}) \coh{X}{pq} (\jones{G}{q} \jones{U}{} )^\herm.
\]

Thus, given a known but unpolarized sky, we can only determine $\jones{G}{p}$ to within an arbitrary unitary ambiguity factor $\jones{U}{}$. In other words, {\em we cannot fix the polarization response of our system without polarized calibrators.} A physical example of such an ambiguity is rotation of all dipoles by the same angle: this cannot be detected through observations of an unpolarized source.

As it turns out, even a polarized calibrator {\em alone} is insufficient, though the matrix algebra gets a bit complicated at this point. The $\coh{B}{}$ matrix is Hermitian positive-definite by construction, and has a Cholesky 
decomposition,\footnote{A Hermitian matrix $\jones{P}{}$ is {\em positive-definite} if $\vec z^\herm\jones{P}{}\vec z > 0$ for all non-zero complex vectors $\vec z$. That $\coh{B}{}$ is positive-definite follows from Sylvester's criterion \citep{Gilbert:SylvestersCriterion}, because $I+Q>0$ and $\mathrm{det}\,\coh{B}{}=I^2-Q^2-U^2-V^2>0$. In fact, the Cholesky decomposition for $\coh{B}{}$ can be worked out directly: $L=\matrixtt{\sqrt{I+Q}}{0}{(U-iV)/\sqrt{I+Q}}{\sqrt{I-Q}}.$} 
i.e. there exists a lower-triangular $\coh{L}{}$ such that $\coh{L}{}\coh{L}{}^H=\coh{B}{}$. For any unitary $\jones{U}{}$, we then have:

\[
(\coh{L}{} \jones{U}{} \coh{L}{}^{-1}) \coh{B}{} (\coh{L}{} \jones{U}{} \coh{L}{}^{-1})^H = 
\coh{L}{} \jones{U}{} (\coh{L}{}^{-1}\coh{L}{})(\coh{L}{}^H\coh{L}{}^{H-1}) \jonesT{U}{}\coh{L}{}^H = \coh{L}{}\coh{L}{}^H = \coh{B}{}.
\]

Therefore, given a single polarized calibrator, we still have an ambiguity factor of $\coh{L}{} \jones{U}{} \coh{L}{}^{-1}$! Physical examples of this are somewhat more elaborate, but perhaps the simplest one is that a source with $Q$ polarization only is insensitive to a certain combination of dipole rotation and gain adjustment. Indeed, for $\coh{B}{}=\matrixtt{I+Q}{0}{0}{I-Q},$ we have $L=\matrixtt{\sqrt{I+Q}}{0}{0}{\sqrt{I-Q}},$ and, given a rotational $\jones{U}{}$, the resulting ambiguity factor is

\[
\coh{L}{} \, \mathrm{Rot}(\phi) \, \coh{L}{}^{-1} = \matrixtt{\cos\phi}{-\sqrt{\frac{I+Q}{I-Q}}\sin\phi}{\sqrt{\frac{I-Q}{I+Q}}\sin\phi}{\cos\phi}.
\]

The upshot of this is that unambiguous calibration of the polarization response of an interferometer requires {\em multiple} polarized calibrator sources, and/or additional assumptions about the sky (e.g. $V=0$, which was a common assumption in the pre-RIME era). \citet{ME5} explores these issues in more detail.

We should note that though the matrix equations above may seem somewhat complicated, they are far more succinct and complete than any scalar equations that have been used to describe polarization calibration prior to the RIME. Once again, the RIME provides a rigorous mathematical language to describe what is otherwise an extremely tricky problem.

%%%
%%% \subsection{\label{sec:lsm}Source and sky models}
%%%
%%% * alternative source representation (shapelets, etc).

\section{\label{sec:ddes}Direction-dependent effects (DDEs)}

Most of the problems associated with \emph{non-trivial} DDEs are already pointed to by Eq.~(\ref{eq:me-allsky}). The fundamental assumption of traditional selfcal is that DDEs are trivial, meaning that:

\begin{itemize}
\item Each observed visibility $\coh{V}{pq}$ is a measurement of the sky coherency function $\coh{X}{}(\vec u)$ at point $\vec u_{pq}$, corrupted by some combination of multiplicative (per-antenna or per-interferometer) gain terms.

\item The coherency function $\coh{X}{}(\vec u)$ is a Fourier transform of the apparent sky $\coh{B}{\mathrm{app}}(\vec l)$ (see also Eq. \ref{eq:me-allsky-simple}).
\end{itemize}

DDEs are a multiplication in the $lm$ plane, which corresponds to a convolution in its Fourier counterpart, the $uv$ plane. That is, in the presence of non-trivial DDEs $\jones{E}{p}(\vec l)$ (including a non-trivial $W_p$ term), the observed visibility is actually a \emph{convolution} of the sky coherency. Assuming $\jones{G}{p}\equiv1$ for the moment, Eq.~(\ref{eq:me-allsky}) then gives us:

\begin{eqnarray}\label{eq:dde-convolution}
\nonumber \coh{V}{pq} & = & \coh{X}{pq}(\vec u_{pq}), \\
\coh{X}{pq} &=& \jones{U}{p} \circ \coh{X}{} \circ \jonesT{U}{q}
\end{eqnarray}

where ``$\circ$'' is a matrix convolution (i.e. following the same rules as matrix multiplication, with each elementary multiplication replaced by a convolution), and the convolution kernels $\jones{U}{p}$ are Fourier transforms of the
sky-Jones terms $\jones{E}{p}$. We can rewrite this equation to emphasize the time variability, and the fact that any given interferometer $pq$ only samples one point $\vec u_{pq}$ of the $uv$ plane at a time:

\begin{eqnarray}\label{eq:dde-convolution-t}
\coh{V}{pq}(t) & = & \coh{X}{pq}[t](\vec u_{pq}(t)), \\
\nonumber \coh{X}{pq}[t] & = & \jones{U}{p}[t] \circ \coh{X}{} \circ \jonesT{U}{q}[t], \\
\nonumber \coh{X}{} & = & {\cal F} \coh{B}{}, \; \jones{U}{p}[t] = {\cal F}\jones{E}{p}[t]
\end{eqnarray}

This equation captures the heart of the DDE problem: DDEs convolve the ``ideal'' visibilities, with (in the general case) a different kernel per every antenna and time sample. Instead of sampling one $uv$ plane ($\coh{X}{}$), we have a separate $uv$ plane per each $pq$ and time interval ($\coh{X}{pq}[t]$), and we're sampling each such plane at only one (or at most a handful) of points. Convolution is not uniquely reversible at the best of times; with such limited sampling it is even less tractable. This is the reason why \emph{in the presence of DDEs, corrected visibilities} (in the sense of Sect.~\ref{sec:implicit-me-newstar}) \emph{do not exist.} To be more precise, they may exist in the mathematical sense, but recovering them is an inverse (and ill-posed) problem.

In this section, I will first consider the two common sources of DDEs: the ionosphere and the primary beam, and
then discuss some proposed methods of dealing with them.

\subsection{$E$-Jones: beam-related DDEs\label{sec:EJones}}

The primary beam gain, commonly designated as the $E$-Jones, is the single most ubiquitous DDE (since every telescope, after all, has a beamshape of some kind), and probably the most problematic.\footnote{In the general formulations above, I used $\jones{E}{}$ to refer to {\em all} DDEs in the signal path. At the risk of confusion, this section will also use  $\jones{E}{}$ for the beam-related Jones term in particular. The ubiquitous nature of beamshapes, and the problems associated with them, is perhaps a justification for using ``E'' as the ``representative'' DDE letter.} The implicit simplifying assumption of 2GC packages is that the interferometer observes an ``apparent sky'': that is, some true sky $\coh{B}{}(l,m)$, attenuated by a {\em power beam} $|E(l,m)|^2$. Given a reasonably accurate model for the beam, the final images can be multiplied by $|E(l,m)|^{-2}$ to correct the flux scale (at the cost of increasing the image noise away from centre).

In RIME terms, this classical assumption corresponds to an $E$-Jones that is a trivial DDE (i.e. constant in time, and same across all stations), but also the same for both receivers and thus scalar: $\jones{E}{p}(t,l,m) \equiv E(l,m)$. We can then commute the $E$ term in the apparent sky equation (\ref{eq:me-allsky}), which becomes
a simple multiplication of the true sky $\coh{B}{}$ by $EE^\herm=|E|^2.$ (Incidentally, this also shows why classical selfcal does not concern itself with the complex phase of the primary beam.) 

Real-life beams deviate from these assumptions in a number of ways, some of them less well understood than others.

\subsubsection{The WSRT and VLA $E$-Jones\label{sec:EJones:wsrt}\label{sec:EJones:vla}}

The WSRT primary beam is commonly approximated as:

\[
E(l,m) = \cos^3(C\nu\sqrt{l^2+m^2}),
\]

where $C$ has a very mild dependence on $\nu$ (i.e. is effectively constant for a given band). This model is only valid for the main lobe, down to about the 10\% level. \citet{Popping-Braun:WSRT-beam} have made a detailed empirical study of the WSRT primary beam, which shows significant four-fold symmetric structure out in the sidelobes (caused by the feed legs). More significantly, they have shown a quasi-periodic ``ripple'' in the off-axis beam gain as a function of frequency, with a period of $\sim17$ MHz. This is commonly seen in the observed spectra of off-axis sources.

Similarly to the WSRT $\cos^3$ model, the VLA primary beam has a reasonable analytic approximation using Jinc functions, which is valid to about the 5\% level of the main lobe \citep{Uson-Cotton:VLA-beam}. \citet{Brisken:VLA-beam} has made electromagnetic simulations that show the sidelobe structure. What significantly complicates the VLA case is {\em beam squint} (the beam pattern of the R and L receptors being offset w.r.t. the pointing centre due to the feeds being off-axis), and parallactic angle rotation.

\subsubsection{Parallactic angle rotation}

An alt-az mount telescope, without a dish derotator such as that designed into ASKAP \citep{ASKAP}, has an intrinsically time-variable beamshape in the $lm$ frame, as the nominal beam pattern rotates with parallactic angle. Like any DDE, this causes significant spatial artefacts around off-axis sources that cannot be addressed by classical selfcal. This has been a serious dynamic range limitation at the VLA, but some recent developments promise to alleviate the problem. \citet{Uson-Cotton:VLA-beam} describe a CLEAN-like algorithm (implemented in the Obit package) that corrects these artefacts during deconvolution; the RIME-derived AW-projection method of \citet{SB:imageplane} can correct them during imaging. Note that both methods rely on an \emph{a priori} beam model, and have, to date, been only been applied to VLA data, for which the Brisken simulations provide a very detailed beam model. It remains to be seen whether the more approximate models available for other instruments will prove to be a limiting factor.

The WSRT's equatorial mounts (and ASKAP's derotator) keep the beamshape stationary in the $lm$ frame, thus avoiding this problem entirely.

A particularly troublesome situation arises when a sufficiently bright source is located in a sidelobe or near a null, where sky rotation causes rapid variation in the beam gain, and the accuracy of existing beam models is low. Such sources have to be calibrated and subtracted separately, either via some kind of peeling procedure, or by using the differential gain approach described in Sect.~\ref{sec:dEs}. Even at the WSRT, where rotation is not an issue and the beam gain remains (at least in principle) constant in time, sources in a sidelobe need to be treated very carefully, due to the rapid spectral variation caused by the 17 MHz ripple.

\subsubsection{Instrumental polarization}

Instrumental polarization comes about due to the beam patterns of the two receptors being non-identical. In RIME terms, this corresponds to $E$-Jones being diagonal rather than simply scalar:

\[
\jones{E}{}(l,m) = \matrixtt{e_x(l,m)}{0}{0}{e_y(l,m)},
\]

which causes an unpolarized off-axis source to ``acquire'' some $Q$ (or $V$, if using circular receptors):

\[
\coh{B}{\mathrm{app}} = \matrixtt{e_x}{0}{0}{e_y} 
\matrixtt{1}{0}{0}{1}
\matrixtt{e_x}{0}{0}{e_y}^\herm =
\matrixtt{|e_x|^2}{0}{0}{|e_y|^2}
\]

The WSRT case is rather simple: the beamshape of each dipole is slightly elongated rather than circularly symmetric. Since these beamshapes are stationary w.r.t. the sky, the net result is an ``apparent sky'' with a non-uniform polarization response: 

\[
\coh{B}{\mathrm{app}} = \matrixtt{|e_x|^2(I+Q)}{e_x e^*_y(U+iV)}{e_x e^*_y(U-iV)}{|e_y|^2(I-Q)}
\]

Similarly to power beam attenuation, this effect can be removed (to the extent that the primary beam is known) via a linear correction to the final images.

For the VLA, non-identical receptor beams are caused by the aforementioned squint; the squint offset rotates with parallactic angle (and thus as a function of time). This leads to a rather complicated picture of instrumental polarization, but is essentially the same problem (with the same solutions) as primary beam rotation.

Note that in contrast to the the WSRT case, the simulations of \citet{Brisken:VLA-beam} show that the VLA $E$-Jones has non-trivial elements on the off-diagonal. This is an example of \emph{direction-dependent polarization leakage}. Leakage has been commonly associated with slight errors in dipole orientation, electromagnetic cross-talk, etc., and treated as a direction-independent effect \citep{ME1,JEN:note185}; Brisken's results demonstrate that it is actually a DDE.

Finally, it should be mentioned that the polarization aberration described by \citet{Carozzi:ME3D} can also be treated as direction-dependent instrumental polarization \citep[see Paper I,][Sect.~5.4]{RRIME1}.

The RIME makes it explicit that effects as (variable) primary beam attenuation, instrumental polarization, and leakage, which are treated separately (if at all) in 2GC, can in fact be represented by a single Jones term, and treated via a single mechanism. Perhaps the most stark example of this is provided by aperture array beams, such as those of LOFAR \citep{Yatawatta:LOFAR-beam}. With the dipoles of an aperture array fixed on the ground, 
$\jones{E}{}(l,m)$ towards any specific sky direction exhibits complex time-dependent behaviour in all four matrix elements. This completely blends the boundary between primary beams, leakage and instrumental polarization.

\subsubsection{Pointing errors \& dish deformation\label{sec:pointing}}

All telescopes mispoint to some extent. This is caused by gravitational load, thermal expansion, wind pressure, errors in the drive mechanics or even the control software, etc. In RIME terms, this can be represented by a station-dependent offset in the beam pattern, causing a nominally identical beamshape $\jones{E}{}$ to produce a different response per station:

\begin{equation}\label{eq:mispointing}
\jones{E}{p}(l,m) = \jones{E}{}(l+\delta l_p,m+\delta m_p)
\end{equation}

The offset $\delta l_p,\delta m_p$ is, in general, time-variable. Since the effect of mispointing on observed visibilities is roughly proportional to $\partial\jones{E}{}/\partial l$ and $\partial \jones{E}{}/\partial m$, it is lowest at the centre of the beam (where the beamshape is flat), and highest on the flank of the beam and around the nulls. Classical selfcal tends to  ``absorb'' the effect of mispointing in the direction of the dominant source
into the per-station amplitude gain solutions. 

Mispointing is thought to be a major source of off-axis errors in WSRT and VLA maps, and thus has been the subject of many studies. \citet{SB:pointing} proposes a modification to the selfcal algorithm called {\em pointing selfcal}, which consists of solving for the $\delta l_p,\delta m_p$ parameters during selfcal. This is predicated on having accurate models for both $\jones{E}{}(l,m)$ and the off-axis sources, and sufficient SNR to constrain the solution. Pointing selfcal has been shown to work with simulated data, and recently with real VLA observations (Bhatnagar, priv. comm.) Paper III \citep{RRIME3} will discuss a different approach to the pointing problem.

The environmental factors responsible for mispointing can also cause deformation of the dish surface. The resulting changes to $\jones{E}{}(l,m)$ are rather more difficult to predict and quantify, and little work has been done on the subject. \citet{Harp:ATA-beams} show significant thermal-related deformations at the Allen Telescope Array (ATA).

\subsection{Ionosphere \& troposphere}

The ionosphere becomes a particularly troublesome DDE at low frequencies, owing to the $\propto\nu^{-1}$ behaviour of ionospheric phase delay, and $\propto\nu^{-2}$ behaviour of Faraday rotation. For a more detailed look at the ionosphere and its effects on signal propagation, see \citet[Sect.~13.3]{tms} and \citet{Intema:SPAM}. Below I will briefly summarize ionospheric effects in terms of the RIME.

\subsubsection{Ionospheric phase}

Ionospheric phase delay is caused by excess pathlength due to refraction. In the RIME formalism, it corresponds to a scalar Jones term: $Z_p=\mathrm{e}^{i\zeta_p}$, where $\zeta_p\propto T \nu^{-1}$, and $T$ is the Total Electron Content along the line-of-sight. Phase delay can easily reach $10^2-10^4\,\mathrm{rad}$ at lower frequencies, with variations on relatively short timescales and small spatial scales, thus making for a rather severe DDE. Following \citet{Lonsdale:4regimes}, we can identify distinct observational regimes based on the size of the array ($A$), the projected size of the FoV ($V$), and the scale structure of the ionosphere ($S$), i.e. the spatial scale on which ionospheric phase is approximately linear. The first criterion is FoV size:

\begin{description}
\item[Narrow FoV:] $V\ll S$, making ionospheric phase effectively constant over the FoV ($Z_p(\vec l)\equiv Z_p(0)$), and thus a DIE.
Since $Z_p$ is scalar, it can be commuted to any position in the RIME and absorbed into another Jones term, such as the per-antenna complex gain that is solved for during regular selfcal.
\item[Wide FoV:] $V\ge S$, and therefore $Z_p$ is properly direction-dependent.
\end{description}

The second criterion is array size:

\begin{description}
\item[Tiny array:] $A\ll S$. Ionospheric phase is constant on scales of $A$, thus $Z_p=Z_q$ for all $p,q$. This makes $Z_p Z^\herm_q=1$, so the interferometer does not ``see'' the ionosphere at all.
\item[Compact array:] $A\approx S$. Ionospheric phase is approximately linear on scales of $A$. Crucially, this means that for any direction $\vec l$ and baseline $pq$, the observed phase difference $\zeta_p-\zeta_q$ is proportional to the projection of the baseline $\vec u$ onto the ionospheric screen, and thus:

\begin{equation}\label{eq:Zjones-shift}
Z_p Z^\herm_q \simeq e^{\eta u_{pq} + \xi v_{pq}}
\end{equation}

\item[Extended array:] $A>S$. Different stations of the array are looking through completely different parts of the ionosphere.
\end{description}

The tiny array case is trivial and not considered further. Lonsdale regimes 1 and 2 correspond to narrow FoVs with compact or extended arrays: these can be dealt with using regular selfcal. In regime 3 (wide FoV, compact array), the ionosphere manifests itself as an apparent ``distortion'' of the field: each source is shifted by its own (time-variable) offset $\eta,\xi$. This can be easily seen by inserting the $Z$-Jones given by Eq.~(\ref{eq:Zjones-shift}) into the full-sky RIME of Eq.~(\ref{eq:me-allsky}), and merging it with the complex exponent. 

Finally, Lonsdale regime 4 corresponds to an extended array and wide FoV. This is the regime in which MWA and LOFAR are expected to operate. $Z_p Z^\herm_q$ then results in a baseline- and direction-dependent phase offset, which causes each source in the field to be ``smeared'' with a different PSF. Selfcal tends to take care of the offset towards the dominant source, thus producing an image which is adequate in the vicinity of the dominant source, but gets increasingly distorted away from it.

\subsubsection{Faraday rotation\label{sec:DFR}\label{sec:FR}}

Faraday rotation (FR) is rotation of the EM field vector that occurs during propagation through a medium of free electrons in the presence of a magnetic field. In RIME terms (and assuming a linear-polarization coordinate basis), the corresponding Jones term is a rotation matrix:

\begin{equation}\label{eq:FR}
\jones{F}{} = \mathrm{Rot}\,\beta = \matrixtt{\cos\beta}{-\sin\beta}{\sin\beta}{\cos\beta}, \;\;
\beta \propto \nu^{-2} \int_\mathrm{LoS} B_{\parallel} n_e ds,
\end{equation}

where ``LoS'' stands for line-of-sight, $B_{\parallel}$ is the component of the magnetic field parallel to the LoS, and $n_e$ is the electron density. In a circular-polarization coordinate basis \citep[see Paper I,][Sect.~6.3]{RRIME1}, $\jones{F}{}$ becomes a differential phase delay of the left- and right-polarized components:

\[
\jones{F}{\odot} = \jones{H}{}\jones{F}{}\jonesinv{H}{} = \matrixtt{\mathrm{e}^{i\beta}}{0}{0}{\mathrm{e}^{-i\beta}}
\]

The obvious observational effect of FR is a frequency-dependent rotation of the angle of polarization. FR associated with the interstellar medium can, for purposes of calibration, be considered an intrinsic property of the sky per se. Because of the $\nu^{-2}$ behaviour, ionospheric FR at higher frequencies is practically negligible. For all these reasons, FR has been an obscure effect, largely ignored outside of the field of polarimetry.

This has changed with the advent of large low-frequency arrays such as LOFAR. In 2010, the first LOFAR long baseline (Effelsberg--Exloo) detected a strange effect: at certain frequencies, an \emph{unpolarized} source was showing significant signal in the $XY/YX$ correlations, and practically none in $XX/YY$ \citep{Wucknitz:DFR}. After considerable excitement, this was linked to \emph{differential FR} (DFR). This effect is an excellent example of the explanatory power of the RIME formalism, so it is worth considering in some detail. At low frequencies, ionospheric FR can be as high as several cycles \citep[e.g. 15 cycles at 100 MHz, see][Sect.~10.3]{tms} so the DFR between two stations of a long baseline can reach significant fractions of a cycle. Consider what happens when an unpolarized 1 Jy point source at phase centre ($K_p\equiv0$) is subject to an FR of $\pi/2[+2\pi n]$ on station $p$, and $0[+2\pi n]$ on station $q$. In the absence of other effects, the measured visibility will be:

\begin{equation}\label{eq:DFR}
\coh{V}{pq} = \jones{F}{p}\coh{B}{}\jonesT{F}{q} = \matrixtt{0}{-1}{1}{0} \matrixtt{1}{0}{0}{1} \matrixtt{1}{0}{0}{1} = \matrixtt{0}{-1}{1}{0},
\end{equation}

or in other words, all the original $I$ flux will be detected as $V$! This clearly shows that DFR is not only a polarimetric concern, but is a mainstream calibration problem.

Perhaps the most striking feature of Eq.~(\ref{eq:DFR}) is how it describes a complicated physical effect with very trivial mathematics. This is a great example of the simplicity brought by the $2\times2$ formalism. Interestingly, this very effect was predicted by \citet{ME1} in the original RIME paper, but (perhaps owing to the relative opacity of the $4\times4$ Mueller formalism, with which it was described) was not immediately recalled when actual DFR was detected\footnote{According to James Anderson (priv. comm.), the VLBI community was aware of the implications of DFR during the 1970s, and this was a major reason for choosing circularly polarized receptors. Recall that in the circular polarization frame, DFR (or indeed any geometric rotation) becomes a simple phase effect, and can be subsumed into the overall phase calibration. I haven't been able to locate a citation for this. There are other compelling reasons for using circular receptors in VLBI: parallactic rotation being easier to deal with is one of them.}.

\subsubsection{Refraction, curvature and absorption}

Ionospheric absorption is a relatively small amplitude effect \citep[e.g. 0.1 dB at 100 MHz and ZA=60$\degr$, see][]{tms}, and is mostly subsumed by the overall gain calibration. Differential absorption makes for a non-trivial DDE, but this is tiny.

Ionospheric refraction causes an apparent shift of position of the source within the primary beam. This can be on the order of $0.05\degr$ (at 100 MHz and ZA=60$\degr$). The corresponding change in primary beam gain can be a significant effect, but is probably not in excess of that caused by uncertainties in the primary beam pattern itself. It can therefore be absorbed by whatever primary beam calibration scheme is applied to the data.

Finally, Anderson (priv. comm.) has pointed out that refraction through a curved ionosphere should produce a phase DDE, due to the fact that the apparent baseline (i.e. the baseline as seen by the refracted wavefront) changes. The Anderson effect should be detectable on LOFAR's long baselines, but it is not yet clear whether it can be separated from $Z$-Jones per se.
 
\subsubsection{The troposphere\label{sec:troposphere}}

The troposphere adds its own phase delay, with a roughly $\propto\nu$ behaviour. Because most of the effect actually happens very close to the ground, tropospheric phase delay $\jones{T}{p}$ is essentially a Regime 2 effect (i.e. a DIE), and can be subsumed into the overall complex gain calibration.\footnote{Because of the $\propto\nu$ behaviour, this is not necessarily true at sub-mm frequencies. The Atacama Large Millimetre Array (ALMA) will rely on water-vapour radiometers for proper tropospheric phase calibration.}

Tropospheric refraction can be significant at low elevations \citep[Sect~10.1]{tms}, so telescopes incorporate a pointing correction to account for it. \emph{Differential} tropospheric refraction (DTR), caused by the curvature of the Earth (i.e. by different antennas ``seeing'' a source at slightly different elevations) should cause a very small DDE. There are hints of this in high-dynamic-range WSRT data (de Bruyn priv. comm.), but more work is required to confirm detection of this. Likewise, an analogue of the Anderson effect should also apply to the troposphere, but it is not clear whether this can be detected.

\subsection{Correcting for known DDEs\label{sec:dde-correction}}

Even when a (non-trivial) DDE is known (whether \emph{a priori} or from calibration), correcting for it is a non-trivial problem. Several approaches to this have been proposed.

\subsubsection{Facet imaging}

If a DDE is known (and constant in time), it may be trivially corrected for in a single direction $\vec l_0$ by applying the inverse of the Jones term $\jones{E}{p}(\vec l_0)$. For example, given the observed visibilities in Eq.~(\ref{eq:me-allsky}), we can apply correction factors of $\jonesinv{E}{p}(\vec l_0)\jonesinv{G}{p}$ and $(\jonesinv{E}{q}(\vec l_0)\jonesinv{G}{q})^{\herm}$. The resulting visibilities will then be given by:

\begin{eqnarray*}
\coh{V}{pq}^{(0)} & = & {\cal F}( \jones{\tilde E}{p} \coh{B}{} \jonesT{\tilde E}{q} ), \\
 & & \mathrm{where} \; \jones{\tilde E}{p}(\vec l) = \jones{E}{p}(\vec l)\jonesinv{E}{p}(\vec l_0).
\end{eqnarray*}

We can then use standard imaging techniques (i.e. the inverse Fourier transform) to compute $\coh{B}{}^{(0)} = {\cal F}^{-1}\coh{V}{}^{(0)}$. Since $\jones{\tilde E}{p}\to1$ with $\vec l \to \vec l_0$, the resulting image is equal to the ``true'' sky at $\vec l_0$  ($\coh{B}{}^{(0)}(\vec l_0) =  \coh{B}{}(\vec l_0)$), and diverges from it as we get away from $\vec l_0$. This is the essence of the \emph{facet} (or \emph{polyhedron}) imaging technique pioneered by Cotton and Schwab \citep[for an overview, see][]{faceting}. The direction $\vec l_0$ corresponds to the center of a facet. By imaging many small facets (each with its own correction factor), and stitching the resulting images together, we can approximate the ``true'' sky to arbitrary precision (by making the facets suitably small). Facet imaging is available in many 2GC packages, and is well-tested and understood. Its major drawback is the high computing cost (when many facets are involved), and the fact that time variability in $\jones{E}{p}$ cannot be taken into account. 

\subsubsection{AW-projection}

A far more promising alternative is suggested by \emph{convolutional function} approaches. The first of these was the \emph{W-projection} algorithm proposed by \citet{Cornwell:wproj}, which corrects for the $W_p$ term on-the-fly during imaging. This is now routinely available in the CASA imager (and also the in {\tt lwimager} tool of the {\tt casarest} package, which shares the same codebase). \citet{SB:imageplane} have generalized this approach to arbitrary DDEs. The resulting \emph{AW-projection} algorithm has been tested in an experimental version of CASA, and it is planned to make it available in future releases (Bhatnagar priv. comm.)

The crucial insight underlying the AW-projection algorithm is that a convolution such as Eq.~(\ref{eq:dde-convolution-t}) can be efficiently computed both in the forward direction, during the degridding step (when predicting visibilities from an image), or in the reverse direction, when gridding visibilities for imaging, on the condition that $\jones{U}{p}$ has limited support (i.e. is significantly non-zero only within a limited area around the origin), which is the same thing as $\jones{E}{p}$ being sufficiently smooth. If we further assume $\jones{E}{p}$ to be (approximately) unitary (i.e. $\jones{E}{p}\jonesT{E}{p}\approx1$), then Eq. ~(\ref{eq:dde-convolution-t}) may even be (approximately) inverted by computing the convolution $\jonesT{U}{p}[t] \circ \coh{V}{pq} \circ \jones{U}{q}[t]$. There is a fixed computational cost associated with the extra convolution kernels, but it scales to wider fields a lot more favourably than the facet imaging approach.

In other words, AW-projection provides an accurate method to apply known DDEs in the forward direction (i.e. when predicting visibilities from a model image), and an approximate method to correct for them in the reverse direction (when imaging).

While W-projection has been in use for a while and is well-tested, the limitations of the more general AW-projection method are still poorly understood. In particular, it is not clear how (or whether) dynamic range is limited by (a) non-unitarity of $\jones{E}{p}$, and (b) the fact that high-order terms in $\jones{U}{p}$ are ignored (i.e. the limited support assumption). No doubt this understanding will improve as implementations of the algorithm become widely available to the community.

\subsubsection{Subtraction in the $uv$-plane\label{sec:subtraction-uv-plane}}

Given a known sky model, the most straightforward way of dealing with a known DDE is to directly evaluate Eq.~(\ref{eq:me-allsky}) in the forward direction, and subtract it from the observed visibilities. This gives us the residuals $\coh{R}{pq} = \coh{D}{pq} - \coh{V}{pq}$, which can then be corrected for the DIEs. Once imaged, they will still be subject to DDEs on the same relative level. However, if a significant portion of the flux is accounted for by the sky model, then the \emph{absolute} level of DDE-related artefacts will be much lower, perhaps even below thermal noise (if the sky model is sufficiently deep -- and a sufficiently deep model is a requirement for calibration anyway). The sky model itself can be added (``restored'') directly into the residual images with no error. This method was used for the reduction of Paper III \citep{RRIME3}, and produced the ``showcase'' image of Fig.~1 therein.

For a sky model composed of discrete source components, this is also called the \emph{DFT approach}, since evaluating Eq.~(\ref{eq:me-allsky}) on a per-source basis is equivalent to doing a brute-strength Discrete Fourier Transform. There has been considerable debate in the literature and at meetings about the relative merits of the DFT approach vs. FFT-based methods such as AW-projection. DFTs have the advantage of maximum precision (at least to the extent that the DDE is known), but are very expensive computationally, since they scale linearly with the number of sources being modelled. AW-projection is approximate (see above), but its computational cost scales much better, as it only depends on resolution.

It should be made clear that the two approaches are complementary rather than mutually exclusive, and can be favourably combined (provided compatible implementations are available, which is a matter of some urgency), by using DFTs for the brighter sources in the field, and AW-projection for the fainter ones. By choosing a flux threshold, one can then achieve a clear trade-off between accuracy (and, ultimately, dynamic range) and computational cost. 

\subsection{Calibrating the unknown DDEs}

\subsubsection{Selfcal contamination}

None of the 2GC packages provide any explicit capabilities for calibration of the unknown DDEs, since they all assume an implicit RIME similar to Eq.~(\ref{eq:newstar-rime}), with a single direction-independent gain term. Consider a very simplified picture, with a field consisting of only two discrete point sources with brightnesses of $\coh{B}{0}$ and $\coh{B}{1}$, and assume DIEs of unity. The observed visibilities $\coh{D}{pq}$ are then given by Eq. (15) of Paper I \citep{RRIME1}, with $\jones{G}{p}\equiv1$:

\begin{eqnarray}\label{eq:selfcal-dpq}
\coh{D}{pq} & = & \jones{E}{0p} \coh{X}{0pq} \jonesT{E}{0q} + \jones{E}{1p} \coh{X}{1pq} \jonesT{E}{1q} + \coh{N}{}, \\
\nonumber && \mathrm{where}\; \coh{X}{spq} = K_{sp} \coh{B}{s} K^\herm_{sq}
\end{eqnarray}  

and $\coh{N}{}$ is a $2\times2$ matrix of Gaussian noise. Traditional selfcal (assuming a perfectly known sky model) then attempts to fit $\coh{D}{pq}$ with the following RIME:

\begin{eqnarray}\label{eq:selfcal-vpq}
\coh{V}{pq} & = & \jones{G}{p} ( \coh{X}{0pq} + \coh{X}{1pq} ) \jonesT{G}{q}
\end{eqnarray}

in a least-squares sense, over all baselines $pq$. Obviously, the best-fitting $\jones{\tilde G}{p}\to\jones{E}{0p}$ as $\coh{B}{1}\to0$. On the other hand, if $\coh{B}{1}\simeq\coh{B}{0}$, $\jones{\tilde G}{p}$ will be some kind of average between $\jones{E}{0p}$ and $\jones{E}{1p}$. Because of the complex phase behaviour in the $K$ terms, this is difficult to analyse in detail. To get a qualitative picture, let us consider the scalar case. Assume that $\jones{E}{p}$ is scalar and purely real, and that the sources are unpolarized, so $\coh{B}{s}$ is scalar as well: $\coh{B}{s}=I_s$. We can see that the biggest discrepancies (in amplitude) occur when the phases of the additive terms in Eq.~(\ref{eq:selfcal-dpq}) add either constructively or destructively. In these two cases, we get:

\begin{eqnarray*}
|\coh{D}{pq}| & = & E_{0p} E_{0q} ( I_0 \pm \frac{E_{1p} E_{1q}}{E_{0p} E_{0q}} I_1 ) \\
|\coh{V}{pq}| & = & |\jones{G}{p}| |\jones{G}{q}| ( I_0 \pm I_1 ) \\
\end{eqnarray*}

For any non-trivial array configuration, each baseline has a different fringe rate, so at any point in time some baselines will be closer to constructive addition, and others will be close to destructive addition. Therefore, no set of
$\jones{\tilde G}{p}$ can achieve a perfect fit of $\coh{D}{pq}$ to $\coh{V}{pq}$. However, from the above we can infer an upper bound on the relative error of the fit:

\begin{eqnarray}\label{eq:contamination}
1 -\Xi_{0,1} \le & { \displaystyle \frac{|\coh{V}{pq}|}{|\coh{D}{pq}|} } & \le 1 + \Xi_{0,1} \\
\nonumber & \Xi_{0,1} & \equiv \max_p \left| \frac{E_{1p}}{E_{0p}} \right|^2\frac{I_1}{I_0}
\end{eqnarray}

I shall call $\Xi_{0,1}$ the \emph{selfcal contamination} factor [of source 1 into source 0]. I do not have a formal proof for a lower boundary on the error terms in Eq.~(\ref{eq:contamination}), but extensive simulations with MeqTrees suggest that it is also proportional to $\Xi_{0,1}$. We can therefore summarize these considerations as follows: in the presence of DDEs, traditional selfcal will tend to subsume the DDEs in the direction of the dominant source into its selfcal gain solutions; the fitted visibilities will be subject to \emph{contamination} from the unmodelled DDEs towards the next-brightest source, with a relative error proportional to $\Xi_{0,1}$.

Similar considerations apply to any discrepancies (i.e. missing sources, etc.) in the sky model. Ultimately, selfcal contamination makes itself felt via artefacts in the resulting images, which can be extraordinarily complicated and counter-intuitive \citep[for an example, see Fig.~17 of Paper III,][]{RRIME3}.

\subsubsection{Peeling\label{sec:peeling}}

The \emph{peeling} algorithm was originally proposed by \citet{JEN:peeling} as a way of calibrating and removing DDEs from bright sources one by one, in order of decreasing brightness. Since its introduction, the term ``peeling'' has been misunderstood and diluted to the point where it is occasionally used to describe {\em any} technique incorporating direction-dependent solutions, but this is incorrect. In its original formulation, peeling refers to a very specific calibration algorithm:

\begin{enumerate}

\item A normal selfcal solution is performed, using an equation such as (\ref{eq:newstar-rime}). The resulting $\jones{\tilde{G}}{p}$ solutions will tend to incorporate DDEs in the direction of the brightest source $s_0$.

\item The prediction for $s_0$ is subtracted from the data. This is the ``peeling'' step per se: our best estimate for the visibility contribution of $s_0$ is, in a sense, peeled away.

\[
\coh{D}{pq}^{(1)} = \coh{D}{pq} - \jones{\tilde{G}}{p} \coh{X}{s_0 pq} \jonesT{\tilde{G}}{q}
\]

\item Optionally, the $\coh{D}{pq}^{(1)}$ visibilities are corrected by applying $\jonesinv{\tilde{G}}{p}$.

\item Optionally, the $\coh{D}{pq}^{(1)}$ visibilities are phase-shifted to the position of the next-brightest source $s_1$ and averaged down in time and frequency (to smear out the contribution of other sources).
 
\item The $\coh{D}{pq}^{(1)}$ visibilities are presumably dominated by source $s_1$. We now go back and repeat the procedure for $s_1$.

\end{enumerate}

Peeling has the considerable advantage that all existing 2GC calibration packages provide sufficient functionality to implement its steps, so it has been widely tested and accepted in the community. 

The major drawback of peeling is that it can be very expensive computationally. Note that the solutions at step 1 are subject to selfcal contamination $\Xi_{s_0,s_1}$. This error is ``frozen in'' at step 2, when the fitted visibilities (for source $s_0$) are subtracted from the data. It can then further contaminate the solutions for $s_1$ (in addition to the contamination $\Xi_{s_1,s_2}$. If the source being peeled is truly dominant, then this contamination can be negligible, but if the brightness of $s_0$ and $s_1$ is comparable, it can become pretty severe. These errors can be driven down by repeated iterations through the peeling cycle (with clever subtraction of sources), at the cost of significant CPU and I/O overhead. This makes peeling impractical when dealing with more than just a few sources. 

\subsubsection{Differential gains\label{sec:dEs}}

The \emph{differential gains} approach is closely related to peeling. It may be thought of as a generalized, simultaneous form of peeling. A detailed practical example will be discussed in Paper III \citep{RRIME3}, but the essence is to use a RIME of the form:

\begin{equation}\label{eq:de}
\coh{V}{pq} = \jones{G}{p} \left( \sum_s \Delta\jones{E}{sp} \coh{X}{spq} \Delta\jonesT{E}{sq} \right)  \jonesT{G}{q},
\end{equation}

and solve for $\jones{G}{p}$ on small time/frequency scales (as per normal selfcal), then \emph{simultaneously} solve for $\Delta\jones{E}{ps}$ on larger time/frequency scales, for a subset of fainter sources. The $\jones{G}{p}$ solutions then subsume all DDEs in the direction of the dominant source, while the $\Delta\jones{E}{ps}$ terms account for the \emph{difference} towards the fainter sources. If some of the DDEs are known \emph{a priori}, suitable terms for them can be inserted into the equation above in addition to $\Delta\jones{E}{ps}$. The differential gain solution will then account only for the remaining unknown DDEs.

Note that solving for $\Delta\jones{E}{}$ on a single off-axis source is equivalent to peeling the dominant source and solving for the off-axis source (with suitable solution intervals chosen for each selfcal step). The $\Delta\jones{E}{}$ approach overcomes a lot of the drawbacks of peeling (contamination of solutions and frozen-in errors, the need for repeated selfcal cycles) by doing a single simultaneous solution in one step.

Differential gains share a common weakness with peeling: that of proliferation of degrees of freedom (DoF's). This is partially mitigated by using larger solution intervals, but it is obvious that we cannot simultaneously solve for $\Delta\jones{E}{ps}$ towards \emph{all} sources in a typical field, since that would be gross over-fitting. (Not to mention the CPU cost of solving for that many parameters simultaneously, which would probably become prohibitive first.) 

\subsubsection{Parametrized models and beacon sources}

The DoF issue can be addressed if the DDE in question can be represented by a parametrized model for $\jones{E}{p}$. We can then solve for the parameters of that model (presumably, few in number), and then correct for the resulting $\jones{E}{p}$ estimate using one of the methods of Sect.~\ref{sec:dde-correction}. 

A number of approaches have shown that this is feasible. For the ionosphere, the \emph{field-based calibration} (FBC) method of \citet{Cotton:FBC} uses the position offsets of sources (in individual snapshot images) to fit a global phase screen over the array. The \emph{source peeling and atmospheric modelling} (SPAM) algorithm of \citet{Intema:SPAM} does a similar fit to phase solutions obtained via peeling (in AIPS). Both methods show how to work around the limitations of 2GC packages: since direct fits to visibilities are impossible in the framework of the latter, especially without a fully-fledged RIME, they rely on standard calibration methods (including peeling), and fit a model to the \emph{results} of calibration. \citet{Hull:ata-beam-fitting} have demonstrated a similar approach for $E$-Jones, using source fluxes to fit the FWHM parameter of the ATA beam. 

Given an explicit RIME, it should be possible to fit parametrized models directly to the observed visibilities. The \emph{minimum ionospheric model} (MIM) approach proposed by Noordam is similar to FBC and SPAM, in that it purports to fit a smooth model for ionospheric phase, but is different in that it uses visibilities (but also other sources of data, such as GPS measurements). This requires a software system where explicit RIMEs may be implemented, and so cannot be adapted to 2GC packages, but it has been demonstrated in the LOFAR BBS system, using a simple linear-slope MIM. The pointing selfcal method \citep{SB:pointing} already mentioned above is an application of the same approach to pointing errors.

All these methods have the common feature of relying on \emph{beacon sources}, that is, having enough sources in the field to constrain the solutions. The availability of a sufficient number of beacons is a crucial question for the calibratability of future instruments. I will return to this in the conclusion to Paper III \citep{RRIME3}, after the results presented therein have been considered.

Note that, just as in the DFT-vs.-FFT debate discussed in Sect.~\ref{sec:subtraction-uv-plane}, there is a related dichotomy between the parametrized model approach, and methods based on direction-dependent solutions (peeling, differential gains). The latter methods {\em require} the use of DFTs at the predict stage, since the FFT approach (AW-projection) cannot be applied without a model of $\jones{E}{p}(\vec l)$ for the entire field. Parametrized models, on the other hand, may be applied both via DFT and FFT. 

Once again, I suggest that the two approaches should be treated as complementary. Looking ahead, the results of Paper III \citep{RRIME3} will show that brighter off-axis sources exhibit all sorts of complicated structure in their $\Delta\jones{E}{p}$ solutions, even in the relatively uncomplicated (i.e. low-DDE) case of WSRT 21 cm observations. It is hard to see how this can be captured by a parametrized DDE model to a precision sufficient for error-free subtraction of such sources. This suggests a similar trade-off in accuracy vs. computing cost as that described in Sect.~\ref{sec:subtraction-uv-plane}, leading to the following hybrid approach for dealing with DDEs:

\begin{enumerate}
\item The unknown DDEs are calibrated for via parametrized model(s), which [hopefully] accounts for the bulk of the effect.
\item In addition, $\Delta\jones{E}{p}$ solutions are obtained for the brighter off-axis sources, to account for any deviations from the sky or DDE models towards those sources.
\item The brightest sources are predicted and subtracted via DFT. 
\item Fainter sources are predicted and subtracted via FFT.
\item The residuals are corrected for during imaging using AW-projection.
\end{enumerate}

Note that the sets of sources involved at steps 2, 3 and 4 are conceptually similar to ``Cat I'' and ``Cat II'' sources proposed for LOFAR calibration \citep{JEN:LOFAR3}, but here I suggest three sets rather than two. The exact partitioning of sources into sets determines the accuracy vs. computing cost trade-off.

\subsubsection{Comparative summary of approaches}

It may be interesting to compare the different approaches to a particular class of DDE, for instance pointing error. Pointing errors introduce an $E$-Jones as given by Eq.~(\ref{eq:mispointing}). To date, three relevant approaches have been proposed: pointing selfcal \citep{SB:pointing}, peeling (Sect.~\ref{sec:peeling}) and differential gains (Sect.~\ref{sec:dEs}). Of these, peeling is by far the best tested, since it is available with all 2GC software packages. Differential gains are available in MeqTrees; pointing selfcal is implemented in an experimental version of CASA (Bhatnagar priv. comm.), but is not publicly available at time of writing. This makes a quantitative comparison impossible, but the algorithms may be compared in principle.

The peeling approach and differential gains are very similar in that they attempt to solve for the same \emph{phenomenological} effect: a direction-dependent complex gain term. In essence, peeling approximates a full-sky RIME as:

\[
\coh{V}{pq} = \jones{G}{1p}(\coh{X}{1pq} + \jones{G}{2p}(\coh{X}{2pq} + \jones{G}{2p}(...)\jonesT{G}{3q})\jonesT{G}{2q})\jonesT{G}{1q},
\]

where $\coh{X}{spq}$ is the model coherency of source $s$ (typically a phase-shifted delta function, for a point source model, but Gaussian sources are also possible in e.g. NEWSTAR). Peeling consists of a least-squares solution for for one set of gains at a time (as in regular selfcal), followed by ``temporary'' subtraction of sources for which a solution has been obtained. Differential gains uses an equation like (\ref{eq:de}). First, a regular selfcal step is done to obtain $\jones{G}{p}$ solutions on short time/frequency scales. This is followed by a simultaneous least-squares solution for all the $\jones{\Delta E}{sp}$ terms, on longer time/frequency scales. 

Peeling is subject to selfcal contamination at each stage of the process, due to the as-yet-unsolved-for contributions of fainter sources. This is especially severe when sources have comparable flux. Differential gains overcomes this by solving for all sources simultaneously. In principle, it should be possible to drive contamination arbitrarily low (and thus achieve the same result as differential gains) via several iterations of the peeling cycle, but this is both labour-intensive, and requires many passes through the data.

Both approaches solve for per-antenna, per-direction gains, while overlooking the fact that physically, these are due to a single per-antenna pointing offset (and thus ignoring Eq.~\ref{eq:mispointing}). Pointing selfcal tries to solve for the true offset itself. In effect, it uses a RIME of the form of Eq.~(\ref{eq:dde-convolution}), where the convolutional terms $\jones{U}{p}={\cal F}\jones{E}{p}$ are the aperture illumination patterns, i.e. the Fourier transforms of the primary beams $\jones{E}{p}$, and $\coh{X}{}$ are the full-sky model coherencies. At the heart of the algorithm is a clever minimization scheme, which essentially decomposes $\jones{U}{p}$ into first- and second-order terms of the pointing offsets $\delta l_p,\delta m_p$. This assumes that the primary beam has a functional form, and that it is (at least to zero-th order) Gaussian. 

The advantage of pointing selfcal is that a single per-antenna pointing offset is obtained, and that the entire model sky (including extended emission!), rather than discrete components, is used to constrain the solution. Peeling and differential gains solve for the total effective gain in each direction, and are less well-constrained by definition. On the other hand, the latter two approaches will happily absorb all unknown DDEs into the direction-dependent solution, while it is yet unclear to what extent pointing selfcal is robust in the presence of other DDEs. The fact that the entire sky is used to constrain the solution also seems to be a double-edged sword. In particular, it is not clear how pointing selfcal is affected by having a bright source near a null or a sidelobe, where the primary beam is particularly poorly approximated by the functional form. 

In terms of performance, pointing selfcal should in principle be the fastest method of the three, since it solves for the least number of unknowns, and also allows for the entire sky to be predicted via an FFT. Differential gains are slower, which is partly due to the use of DFTs for source prediction, although the true bottleneck is the far larger number of unknowns. Peeling, on the other hand, is I/O-bound due to the large number of data passes, which will usually make it the slowest of the lot. 

\section{Conclusions}
\EDIT{
Several authors have developed approaches to the DDE problem based on the RIME, using different (but mathematically equivalent) versions of the formalism. This paper has attempted to reformulate these using one consistent $2\times2$ formalism, and consider how these methods may be combined. 

A look at such DDEs as instrumental polarization (Sect.~\ref{sec:EJones}) and differential Faraday rotation (Sect.~\ref{sec:DFR}) suggests that the study of polarized signals is no longer a side issue of interest only to polarimetry per se. Proper calibration of the new crop of instruments requires that a full-polarization picture be considered from the beginning. Fortunately, the RIME provides just such a picture, by recasting the signal in terms of $2\times2$ coherency matrices rather than $IQUV$ vectors. This allows complicated propagation effects to be described in terms of rigorous and straightforward matrix algebra, and builds valuable links between one's physical and mathematical intuition. 
}

\bibliographystyle{aa}

\bibliography{16082}

\end{document}